\documentclass[journal=ancac3,manuscript=article]{achemso}
\usepackage[version=3]{mhchem} 
\usepackage{graphicx}
\usepackage{graphics}
\usepackage{color}
\usepackage{amsmath}
\usepackage{lmodern} 

\author{M. Pilo-Pais}
\email{mgp8@phy.duke.edu}
\author{A. Watson}
\affiliation[Duke University]
{$\dagger$Department of Physics, Duke University, Durham NC}
\author{S. Demers}
\affiliation[Duke University]
{$\dagger$Department of Physics, Duke University, Durham NC}
\author{T. H. LaBean}
\affiliation[North Carolina State University]
{$\ddagger$Department of Materials Science and Engineering, North Carolina State University, Raleigh NC}
\author{G. Finkelstein}
\affiliation[Duke University]
{$\dagger$Department of Physics, Duke University, Durham NC}
\email{gleb@phy.duke.edu}
\title{SERS Plasmonic Enhancement using DNA Origami-based Complex Metallic Nanostructures}
\begin{document}
\begin{abstract}
DNA origami is a novel self-assembly technique allowing one to form various 2D shapes and position matter with nanometer accuracy. We use DNA origami templates to engineer Surface Enhanced Raman Scattering (SERS) substrates. Specifically, gold nanoparticles were selectively placed on the corners of rectangular origami and subsequently enlarged via solution-based metal deposition. The resulting assemblies exhibit ``hot spots'' of enhanced electromagnetic field between the nanoparticles. We observed a significant Raman signal enhancement from molecules covalently attached to the assemblies, as compared to control nanoparticle samples which lack inter-particle hot spots. Furthermore, Raman molecules are used to map out the hot spots' distribution, as they are burned when experiencing a threshold electric field. Our method opens up the prospects of using DNA origami to rationally engineer and assemble plasmonic structures for molecular spectroscopy.
\end{abstract}
%

DNA origami is a product of a one-pot reaction in which DNA strands of specific sequences self-assemble into a large structure ($\sim$100 nm) of a predetermined shape,~\cite{Rothemund2006} thereby providing an alternative to conventional top-down fabrication methods. The resulting templates are highly addressable and versatile tools for site-specific placement of various nanocomponents, such as metallic nanoparticles,~\cite{Ding2010a,Hung2010a,Pal2010a,Pal2011,Pilo-Pais2011g} quantum dots,~\cite{Bui2010} fluorophores,~\cite{Steinhauer2009} and carbon nanotubes.~\cite{Maune2010a} It has also been shown that origami templates can serve as platforms for DNA motors,~\cite{Venkataraman2007} DNA walkers,~\cite{Gu2010, Lund2010, Wickham2011} and chemical reactions.~\cite{Voigt2010} Most recently, origami templates have been used to promote interactions between attached nanocomponents, such as plasmonic coupling between gold nanorods~\cite{Pal2010a} or gold nanoparticles,~\cite{Kuzyk2012} as well as enhancement and quenching of fluorophores~\cite{Acuna2012, Acuna2012a} or CdSe QDs~\cite{Ko2013}.

Building on the massively parallel formation of DNA origami and their capability to serve as a nanobreadboard, one can further envision using them as biosensors. One particularly attractive goal is to facilitate Raman spectroscopy, which provides a highly specific chemical fingerprint. Unfortunately, the Raman scattering cross section is small; Surface Enhanced Raman Spectroscopy (SERS) greatly increases the signal by utilizing the regions of intense electric field created near granular metallic surfaces. These ``hot spots'' can be understood as resulting from localized surface plasmon modes resonantly excited by the incident laser. The analyte molecules that happen to be positioned in the hot spots provide disproportionately high contribution to the Raman scattering, resulting in a signal enhancement that is many orders of magnitude.~\cite{Ru2008a, Kleinman2013}

Pairwise complementary DNA strands can be used to position Raman-active molecules between functionalized gold nanoparticles (AuNPs), thus fabricating plasmonic structures with active hot spots.~\cite{Cao2002, Graham2008, Lim2010, Lim2011}. In this paper, we utilized DNA origami to fabricate more complex multi-particle assemblies, and determined their performance as SERS substrates. 


\begin{figure}
\includegraphics[width=1\textwidth, viewport = 60 340 560 560, clip=true]{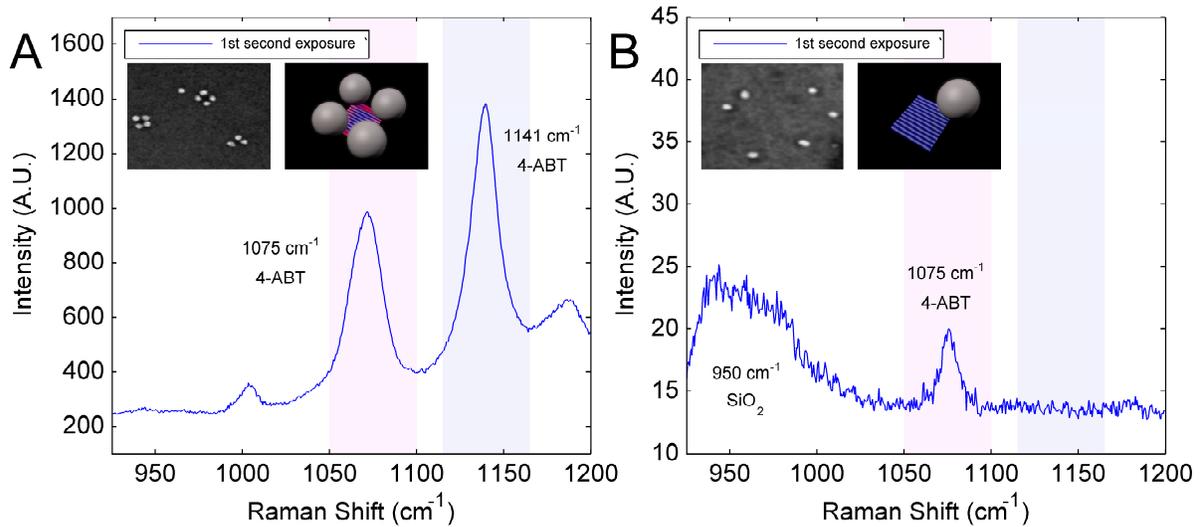}
\caption{Typical SERS spectra of 4-aminobenzenethiol (4-ABT) attached to metal nanoparticles assembled on DNA origami. \textbf{a)} Four-particle assemblies (``tetramers'') which have interparticle hot spots; \textbf{b)} control sample with one AuNP per origami (``monomers'') which lack the interparticle hot spots. Spectra correspond to the first 1 second of the laser exposure and are normalized to the average density of nanoparticles in the illuminated region. Insets: SEM images taken from the measurement areas and cartoons representing the target structure. The origami template is shown in blue, while the red tint in (a) indicates the regions of enhanced electric field (hot spots). Larger area images and higher-magnification views of the individual structures can be found in the supplementary information.}
\end{figure}\label{fig1}

Specifically, we used DNA origami to organize the metallic structures, and then covalently attached Raman-active molecules to the metal. We found that the substrates with four nanoparticles (NPs) per origami produce a strongly enhanced Raman signal compared to the control samples with only one nanoparticle per origami. Indeed, the small gaps between closely spaced nanoparticles result in hot spots, which are absent in samples with individual nanoparticles (\ref{fig1}Figure 1). Furthermore, the Raman signal systematically decayed as a function of the laser exposure time in the samples with four particles per origami. We attribute this behavior to molecular damage caused  by the high electric field at the hot spots. The one-particle control samples lacking the interparticle hot spots exhibited no such decay.

\section{Results and discussion} 

We use DNA origami to control the composition, shape, geometry and arrangement of metallic structures, which in turn determine the local distribution of electromagnetic fields. DNA-metallic assemblies were prepared as we previously reported.~\cite{Pilo-Pais2011g}  Briefly, select staple strands of the standard ``tall'' rectangular DNA origami ($\sim$90x70 $nm^2$) are extended by a specific ss-DNA sequence, referred to as X$_{24}$. The sequence serves as an anchor for AuNPs; to increase the binding probability the anchors were positioned in pairs on adjacent staples. The AuNPs are conjugated with $\sim$5 complementary sequence strands (X$_{24,comp}$) through standard thiol chemistry~\cite{Loweth1999} (Please refer to the Methods section for complete experimental details and to the Supplementary Information section for the list of sequences used on the modified DNA strands).

Two different types of SERS samples were prepared: in the sample with engineered hot spots, four AuNPs were attached to each of the four corners of the origami template (tetramer samples); in the control samples, only one AuNP was placed in one of the corners (monomer samples). Each of the modified DNA template designs was attached to RCA cleaned (SC-1 and SC-2) and oxygen plasma ashed (SPI Plasma Prep II, 20 min, 100 mA) silicon dioxide substrates (1 $\mu$m oxide, Silicon Quest) using a 10x TAE/125 mM Mg$^{2+}$ solution (the final DNA origami concentration was 0.5 nM). Functionalized AuNPs were then added to the solution (final concentration of 10 nM) resulting on the attachment to the origami templates. They were then incubated for 15 min and rinsed in DI water for 15 seconds, followed by gentle drying with nitrogen. We chose to assemble tetramers as opposed to dimers because they exhibit interesting plasmonic properties such as Fano resonances,\cite{Shafiei2013} which have been shown to provide much greater SERS enhancement due to near-field intensity variations.\cite{Ye2012} The present work is a proof of concept that DNA origami can be used to assemble complex metallic structures for optical applications such as SERS substrates.

The structures were then enlarged using a commercial silver metallization kit following the manufacturer directions (HQ silver enhancement, Nanoprobes INC), as performed in our previous works.~\cite{Park2006, Pilo-Pais2011g} The incubation time was 12 min for the tetramer structures. This was the necessary time to have the nanoparticles closely spaced but typically not touching, permitting the formation of interparticle hot spots. The control samples were incubated for 10 min to achieve the same average NP size of ~50 nm. Although measuring the gap between nanoparticles with nanometer precision is challenging, we estimate the average gap size to be below 3 nm (Figure SI2). This separation distance was clearly small enough to produce a strong signal enhancement, as demonstrated from the Raman spectra (Figure 1). We lithographically patterned the samples with a set of markers, which allowed us to identify specific regions where the spectra were taken. Specifically, we spin-coated PMMA-A4 (MicroChem Corp.), followed by a 5 min UV exposure and development to open $\sim$10 $\mu$m windows. The samples were then imaged using a Scanning Electron Microscope (SEM); we selected for further study the windows which showed no abnormally assembled structures such as the multi-particle clusters seen in Figure S2. We also measured the average nanoparticle density in each window, used for normalization purposes. The Raman spectra exhibit no discernible difference before or after lithographic patterning.

\begin{figure}
\includegraphics[width=1\textwidth, viewport= 70 340 550 550, clip=true]{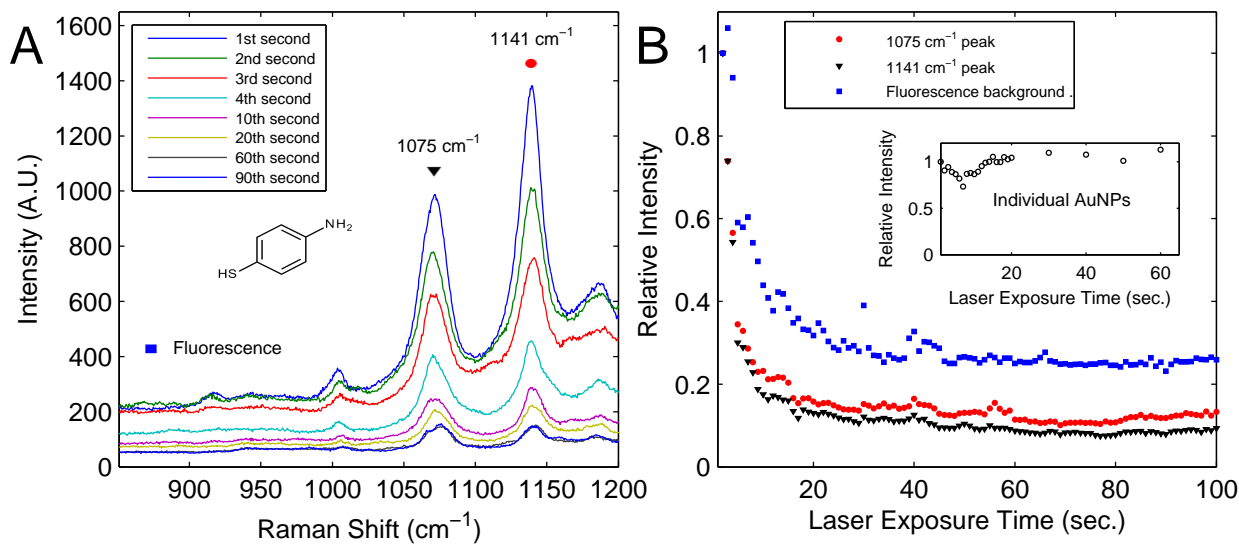}
\caption{\textbf{a)} Raman spectra taken by repeatedly exposing the tetramer sample to 1 second HeNe laser pulses. \textbf{b)} Intensity of the 1075 cm$^{-1}$ and 1141 cm$^{-1}$  Raman peaks (background subtracted) and the fluorescence background as a function of the laser exposure time. The rapid decay is attributed to the photo-damage of the molecules caused by the intense field at the hot spots.} 
\end{figure}\label{fig2}

The samples were then incubated in a 5 mM solution of 4-aminobenzenethiol (4-ABT) in ethanol for two hours (long enough to reach full surface coverage). Placement of multiple Raman molecules throughout the hot spot is imperative to obtain an estimate of the enhancement factor, which is not possible with a single molecule. The thiol functional group ensures that the Raman-active molecules are covalently attached only to the metallic surfaces. The samples were then thoroughly rinsed in pure ethanol to remove any physisorbed molecules. Similarly treated SiO$_2$ substrates without NPs showed no detectable 4-ABT Raman signal, indicating the effectiveness of the rinsing procedure (Incidentally, we did not observe Raman signatures of any other molecules, such as DNA).

The Raman spectra were obtained using a Jobin Yvon LabRam ARAMIS (Horiba, Ltd) spectrometer using a 632.8 nm, 5 mW HeNe laser excitation, focused by a 100x objective to a $\sim$1 $\mu$m spot. \ref{fig1}Figure 1 compares the Raman signal measured from tetramer and monomer samples during the first second exposure at maximal laser intensity ($I_o$). The spectra were normalized to the average NP number in the corresponding lithographic window. We verified that the magnitude of Raman signal per particle was reproducible in other regions. The average relative Raman enhancement per NP is $\sim 100$ in tetramer \emph{vs.} monomer samples. Note that using the monomer samples as a control ensures that the surface concentration of the covalently attached 4-ABT layer is the same as in the tetramer samples. This eliminates the uncertainty in determining the SERS enhancement factor, common for the measurements which use molecular solutions as a control.~\cite{Ru2008a}.

The observed SERS enhancement is naturally explained by the hot spots created in the tetramer sample. Numerical simulations (not shown) indicate that the hot spots are located between pairs of NPs (see cartoon in Figure 3C) similar to the hot spots created in nanoparticle dimers. Unlike dimers, where the enhancement disappears for an electric field perpendicular to the dimer axis, the hot spots in the tetramers should be activated by any laser polarization. Our control monomer samples lack the inter-particle hot spots; although the electric field is also enhanced at the particle poles, the enhancement factor should be much smaller, and we disregard it in the following discussion. Notice also that the 1141 $cm^{-1}$ Raman mode (blue shadow on Figure 1) is no longer observed from the monomer samples. This can be attributed to a strong chemical dependence of the Raman mode, which requires a minimum excitation energy in order to promote a charge transfer.\cite{Kim2011b,Kim2012d}

The importance of the hot spots is further evidenced by the time evolution of the Raman signal. \ref{fig2}Figure 2A shows the sequence of spectra taken during successive 1 second exposures. The signal intensity initially drops rapidly and then saturates. Similar intensity decay has been attributed to photo-damaging of Raman molecules by the enhanced field at the hot spots.~\cite{Etchegoin2007a,Fang2008,Yang2013}. Alternatively, the decay has been assigned to morphological changes of the metallic structures due to heating.~\cite{Soudamini2009} Although it is difficult to identify the mechanism responsible for the decay we observe~\cite{Ru2008a}, we tentatively attribute it to molecular photo-damage. SEM images of the structures before and after the Raman measurement do not show noticeable differences, within a resolution of a few nanometers. It is also important to emphasize that the molecules are covalently attached to the silver particles through the thiol crosslinker, preventing them from leaving the hot spots (photo-desorption).

\begin{figure}
  \includegraphics[width=1\linewidth,viewport = 60 330 500 650, clip=true]{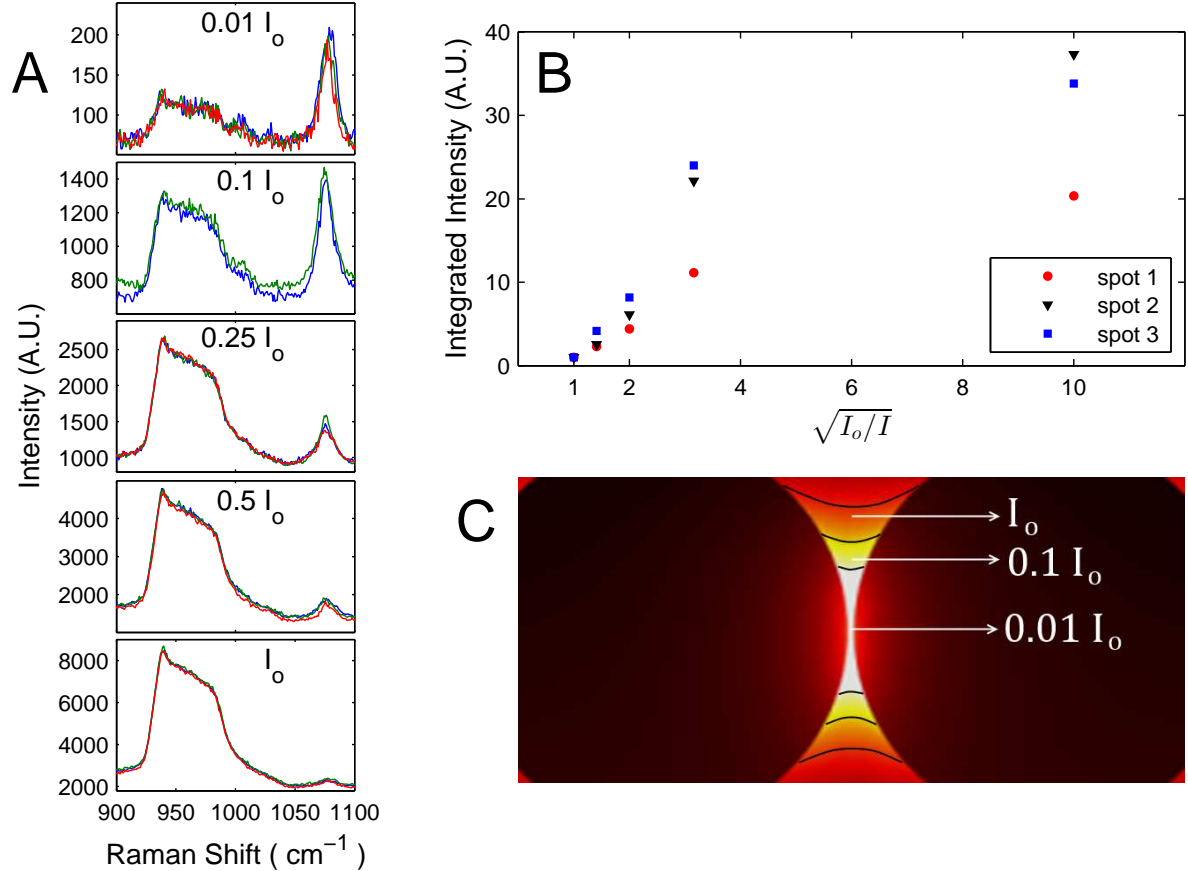}
\caption{\textbf{a)} Raman spectra in the vicinity of the 950 cm$^{-1}$ substrate band and the 1075 cm$^{-1}$ 4-ABT peak. Each spectrum is taken after increasing the laser intensity to 1\%, 10\%, 25\%, 50\% and 100\% of $I_0$ and waiting for the signal to saturate at the new intensity level. Two or three spectra are consecutively taken at each intensity level, demonstrating the saturation of the signal. \textbf{b)} Ratio of the time-saturated 4-ABT Raman peak (background subtracted) to the laser intensity $I$ as a function of $\sqrt{(I_0/I)} \propto 1/E$. The colored data sets correspond to different locations on the sample and are scaled to 1 at $I=I_0$. \textbf{(c)} Cartoon showing the electric field enhancement regions within the hot spot. Optical filters are interchanged in order to increase the incident field in steps, burning the molecules located in progressively larger regions of the hot spots.}
\label{fig3}
\end{figure}

Figure 2B shows the integrated Raman signal for the 1075 cm$^{-1}$ (CS stretch, $7a_{1}$) and the 1141 cm$^{-1}$ (CH bend, $b_{2}$) peaks, as well as the fluorescence background as a function of the laser exposure time. All the signals are normalized to the values measured at the first 1-second exposure. The signal decays rapidly at first, and then saturates at a constant value. The inset of Figure 2B shows the integrated intensity of the 1075 cm$^{-1}$ peak measured from the control monomer sample; no signal decay is observed in this case due to the lack of interparticle hot spots.



The saturation of the Raman signal after about 100 seconds allows us to further characterize the electric field enhancement in the hot spots.~\cite{Etchegoin2007a, Fang2008} We assume that only the molecules experiencing an electric field exceeding a certain critical value are destroyed.~\cite{Ru2008a} No signal decay is observed at 1 \% of the maximal laser intensity $I_0$, indicating that the critical field is not yet reached even in the hot spots. To study the successive photo-damage of the hot spots by the laser, we increased the illumination intensity to 10, 25, 50 and 100 \% of $I_0$. Each successive increase of the intensity is followed by the gradual decay of the Raman signal, similar to the decay shown in Figure 2. This behavior indicates the step-wise expansion of the regions where the field exceeds the critical value and the molecules are photo-damaged (see schematics in Figure 3c). As a result, the ratio of the time-saturated Raman signal to the laser intensity decreases with increasing intensity. 

Figure 3a illustrates this behavior by showing the time-saturated spectra for each subsequent intensity increment, all measured from the same spot. Notice the decay of the relative intensity of the 4-ABT Raman signal at 1075 cm$^{-1}$ as compared to the substrate Raman band centered at 950 cm$^{-1}$. The latter signal was verified to scale proportionally to the laser intensity, while the 4-ABT signal is clearly sub-linear: for example, at the full intensity of $I_0$ the 4-ABT Raman peak is barely visible relative to the 950 cm$^{-1}$ band, while at 1\% of $I_0$ the 4-ABT Raman peak dominates (this figure is not normalized to the number of particles. Also, it is measured from a more dilute sample compared to Figures 1 and 2, which increases the relative prominence of the 950 cm$^{-1}$ substrate band and allows us to visualize the effect). 

In Figure 3b, we plot the ratio of the 4-ABT Raman signal to the laser intensity, $I$. The horizontal axis is $\sqrt{(I_0/I)}$, proportional to the inverse incident electric field.~\cite{Fang2008,Yang2013} Data taken at three different spots are represented by different colors, all normalized to 1 at $I=I_0$. In each of the curves, the relative Raman signal at $I = 0.01 I_0$ is 20-40 times stronger than the signal at $I_0$. In other words, at the maximal laser intensity the high fields of the hot spots damage the molecules responsible for most of the Raman enhancement that was achieved at 1\% of the intensity.  Still, we recall that at $I_0$ the signal measured from the tetramer sample is $\sim$ 100 times stronger compared to the monomers. Since the threshold photo-damage field is not yet reached in the monomer samples, even at $I_0$, their Raman signal should scale proportionally to the laser intensity. Therefore, at 1\% of $I_0$, the enhancement factor of the tetramer could potentially reach 2000-4000. Unfortunately, the direct comparison was not feasible, because the Raman signal of the monomer sample was too weak at $0.01 I_0$, while at $I_0$ the Raman signal from the tetramer samples experiences a very significant degradation prior to the completion of the first 1-second exposure.

\section{Conclusion}

We have shown that DNA origami can be successfully used to engineer substrates for Surface Enhanced Raman Spectroscopy. The Raman signal of 4-ABT molecules deposited on the tetramer NP assemblies is enhanced at least a hundred times (and potentially several thousand) per nanoparticle as compared to control samples with individual nanoparticles. The enhancement is due to hot spots, whose existence was verified by time and intensity-dependent measurements. Our results demonstrate the design capabilities that origami-based metallic structures can offer for spectroscopic and plasmonic applications. In the future, we plan to fully use the addressability of the DNA origami and to custom tune the plasmon resonance frequency of the DNA-metallic structures by exploring different shapes and materials. This will allow us to tailor the plasmonic resonances to match the laser frequency and/or the optimal excitation band of a given molecule. The methodology presented here opens up new possibilities to rationally engineer substrates using DNA origami for optical and plasmonic applications.


\section{Methods}
\small{
\textbf{DNA templates synthesis:}
All DNA sequences were purchased from IDT (Integrated DNA Technologies, Inc.). The modified tall rectangular DNA origami templates were formed using Rothemund design~\cite{Rothemund2006} but with the following modifications: All of the side staples were left out to prevent stacking between multiple origamis. The binding sites were the AuNPs anchor were made by extending two consecutive staples with a short spacer sequence T$_{5}$ followed by a 24 bp DNA sequence (X$_{24}$) (See supplementary information for a list of DNA sequences). The prepared origami ($\sim$5 nM) was filtered from the excess staples by centrifuging using a ultrafiltration centrifuge unit (100 KDa MWCO, Millipore) with three washes of 1x TAE, 12.5 mM Mg$^{2+}$ buffer.

\noindent \textbf{AuNP DNA conjugation:}
Gold Nanoparticles were concentrated and functionalized using a phosphine recipe originally developed by ref ~\cite{Loweth1999} but with some changes: 10 mL of 5 nm, 80 nM AuNPs solution (British Biocell International) were incubated overnight with 3 mg of bis(p-sulfonatophenyl)phenylphosphine (BSPP, Sigma-Aldrich). The AuNPs were concentrated by adding 250 mg of NaCl and centrifuging for 30 min at 800 g. The supernatant was removed without disturbing the AuNP pellet and resuspended on 200 $\mu$L of methanol and 200 $\mu$L of BSPP solution (3 mg in 10 mL DI water). The solution was once again centrifuged for 30 min and its supernatant removed. The AuNPs were resuspended with 200 $\mu$L of the same BSPP solution and incubated for 48 hrs with disulfide-modified X$_{24}$  DNA sequence at a ratio of 1:5 Au:DNA and adjusted to 1x TAE, 50 mM NaCl. Thiolated T$_{5}$ strands were added at a Au:T$_{5}$ ratio of 1:60 to fully backfill the AuNP-DNA conjugates in order to prevent aggregation in a 125 nM Mg environment. Excess DNA strands were removed by running the AuNP-DNA conjugates on a 3\% agarose gel (0.5x TAE) for 25 min, at 10 V/cm. As indicated in ref,~\cite{Pilo-Pais2011g} we find this purification step critical to obtain a high-yield binding. The AuNP-conjugate was recovered using Freeze and Squeeze (Bio-Rad Laboratories) with a typical recovery concentration of 500 nM. 

\noindent \textbf{DNA-Metallic structures and individual nanoparticles formation:} The modified DNA templates were attached to previously RCA cleaned and oxygen plasma ashed (SPI Plasma Prep II, 20 min, 100 mA) silicon dioxide substrates (1 $\mu$m oxide, Silicon Quest) using 10x TAE/125 mM Mg$^{2+}$  solution. The final DNA origami concentration was 250 pM. Functionalized gold nanoparticles (AuNPs) were attached onto the templates by adding 1 $\mu$L of a concentrated solution, to a final concentration of 3 nM and incubated for 15 min and rinsed in water for 5 seconds, followed by nitrogen blow. The structures were then enlarged in size using in-solution silver metallization as indicated by the manufacturer (HQ silver enhancement, Nanoprobes INC) for 12 min for the DNA-metallic structures and 10 min for the individual nanoparticles.

\noindent \textbf{Lithography:} Samples were spin coated with PMMA-A4 (MicroChem Corp) and baked on a hot plate (120 C surface temperature) for 2 min. A copper grid of 2000 mesh (Structure Probe, Inc) was placed on top and the sample was exposed to UV (500 W) for 5 min.

\noindent \textbf{Raman measurements:} 
DNA-metallic structures were incubated on a 5 mM 4-aminobenzenethiol ethanolic solution for 4 hrs followed by a thorough rinse on pure ethanol. Raman spectra were obtained using a Jobin Yvon LabRam ARAMIS (Horiba, Ltd) Raman instrumentation. Measurements were taking using a 5 mW HeNe 632.8 nm laser using 1 second intervals. A 100x objective, resolution grating of 1800 grooves, and a slit of 100 $\mu$m were used on all measurements. The spectra ranged from 1000 to 1600 cm$^{-1}$.
}

\bibliographystyle{unsrt}
\bibliography{library_final}

\begin{acknowledgement}
The authors thank Henry Everitt for his useful suggestions. M. Pilo-Pais thanks Jack Mock and Christos Argyropoulos for valuable discussions on optical measurements. This work has been supported by NSF-ECCS-12-32239.
\end{acknowledgement}

\section{Author contributions}
M.P., T.H.L., and G.F. designed the experiment. M.P., A.W., and S.D. fabricated the samples and conducted the experiment. M.P., A.W., and G.F. analyzed and interpreted the data.
\section{Additional information}
Supplementary Information accompanies this paper

\end{document}